# Using laser and ultrasound devices for breast cancer treatment: studying the effects of gold nanoparticles injection

Heliya sadat Kazemi Siyanaki[1], Fatemeh Rezaei[1], Saeedeh Kabiri[2], Behnam Ashrafkhani[3]

1. Department of Physics, K. N. Toosi University of Technology, Shariati, Tehran 15875□4416, Iran.

2. Department of Physics, Sharif University of technology, Tehran 11155□9161, Iran.

3. Department of Physics, University of Calgary, Canada, Calgary, Alberta

## Abstract

Breast cancer is a disease in which cells in the breast grow out of control. After surgery, chemotherapy, and other invasive treatments, hyperthermia is a suitable choice with the minimal side effects. In this paper, the treatment of breast cancer using a combination of laser and ultrasound irradiation, in the presence of gold nanoparticles, is investigated. In the simulations, the breast tissue is represented as a multilayer structure and the tumor is supposed to consist two parts: a superficial section and a deeper region. In the initial stage, the superficial parts of the tumor, which also contain gold nanoparticles, are exposed to a continuous laser for 50 seconds followed by a cooling period of 20 seconds. Then, for deeper sections, ultrasound irradiation is utilized. The results indicated that the tissue necrosis volume was enhanced by the application of nanoparticles. Furthermore, it was demonstrated that the combinational application of both of laser and ultrasound irradiation could eradicate both the superficial parts of the tumor and the deep parts.

*Corresponding author*: fatemehrezaei@kntu.ac.ir

## 1. Introduction

Cancer is a cellular-level disease that originates from the uncontrolled growth of abnormal cells and can spread to other areas of the body[1]. Among all types of cancers, breast cancer is the most common and the most dangerous cancer in women worldwide[2]. With a global mortality rate of 1 per 5,000 patients, breast cancer has a relatively high mortality rate[3]. There are numerous factors that contribute to the development of breast cancer.  It is believed to be the result of the interaction of genetic factors with environmental factors[4]. There are different methods to treat breast cancer such as surgery[5], radiation therapy[6], hormone therapy[7], ultrasound irradiation[8], and chemotherapy[9]. Generally, for a malignant tumor that spreads outside the breast and lymph nodes, hormone therapy and chemotherapy are preferable choices[3]. In addition, there are two common methods of surgery, mastectomy, in which the entire breast is removed, and lumpectomy, in which only some parts of the breast are removed, however, the mentioned methods are invasive and have side effects such as hair loss, weight gain, menopause, osteoporosis and other complications[3]. Among all common non-invasive methods, laser therapy and ultrasound radiation, which are both based on hyperthermia, have the fewest side effects and are often recommended[10,11]. In hyperthermia treatment, the temperature is increased up to 40 degrees Celsius, which causes tumor necrosis with minimal damage to healthy tissue[12]. The main challenge of hyperthermia method is to create a temperature above 40 degrees Celsius and focus it on the tumor. In recent years, nanoparticles have been proposed for this purpose. As reported in a lot of literatures, the addition of nanoparticles significantly improves the treatment of various cancers. It should be noted that the use of nanoparticles has a considerable effect on destroying tumor cells and reducing the damage to surrounding tissues[13,14].

Several studies have been conducted on the treatment of breast cancer with hyperthermia. For instance, X. Ma et al.[15] used $Fe_3O_4$-Pd Janus nanoparticles for efficient cancer therapy. They showed that $Fe_3O_4$-Pd Janus nanoparticles demonstrated significantly high magnetic-photo heating efficiency and enhanced ROS generation, enabling effective tumor regression. They also presented dual-modal MRI/PA imaging signal enhancement, allowing precise imaging-guided therapy with high spatial resolution and accuracy. Furthermore, Burke et al.[16] examined the combination of multiwalled carbon nanotubes (MWCNTs) and near-infrared (NIR) laser irradiation. Their research showed rapid tumor regression and long-term survival in a mouse model, providing a promising treatment strategy for breast cancer. In addition, Sharma. D et al.[17] investigated the effect of varying hyperthermia and ultrasound-stimulated microbubble (USMB) treatment parameters on tumor cell death and vascular disruption in breast cancer therapy. They demonstrated that combining USMB with hyperthermia enhanced tumor cell death and reduced vascular density in breast tumor xenografts. They also highlighted the role of prolonged ultrasound pulses and increased ultrasound exposure in enhancing particle penetration and tumor response. Hassan et al.[18] performed numerical simulations and analytical calculations to compare the interaction of therapeutic ultrasound with a multilayer tissue at different frequencies and stimulation times, by focusing on calculating the temperature in different tissues. According to their research, the highest temperature observed in the tumor was 45.5°C at a frequency of 1.5 MHz, which was favorable in destroying the cancerous cells without causing significant damage to surrounding tissues. Analytical results represented that the highest temperature predicted in the tumor was 47.77°C, and the two solutions were considered to be in good agreement.

It has also been reported that the integration of different therapeutic methods can produce promising results in the treatment of tumor cells. In 2022, Kabiri and Rezaei[19] examined the effectiveness of combining nanoparticle-mediated laser emission and microwave irradiation in treating liver cancer and showed that integrating these two modalities could considerably enhance the efficiency and accuracy of each of methods. In the present study, the synergistic effect of combining laser photothermal therapy with ultrasound irradiation is investigated through FEM simulation and it is shown that this intelligent integration of methods can help the necrosis of tumors with complex morphology.

## 2. Materials and methods

### 2.1 breast cancer model

In this research, the breast is modeled as a multi-layered hemisphere[3] including epidermis, papillary dermis, reticular dermis, fat and gland containing a tumor. The radius of the breast is assumed to be 65 mm. Moreover, the tumor is assumed to have an arbitrary shape including superficial and deep sections. The tumor's width varies from zero to 14.4 mm with the depth of 15 mm. Figure 1 shows a schematics diagram of the tissue and the tumor designed for the simulation. Tables 1 and 2 show the physical properties of tissue, tumor and blood.

In the present simulation, finite element method (FEM) solves all of the equations and due to the luge number of elements in 3 dimensions and limitations of the solver for calculating this models, the axial symmetry is considered for this geometry. In this paper, ultrasound beam is produced by an acoustic transducer. Figure 2 shows the geometry of the tissue under the acoustic irradiation. It should be noted that in figure 2, both the tissue and the acoustic transducer are immersed in water. The transducer is a bowl-shaped element with a focal length of 62.64 mm, including an aperture with radius of 35 mm, and a hole with the radius of 10 mm in the center. The frequency of the transducer is 1 MHz[20].

In this simulation, three lasers with the similar wavelength of 796 nm and various laser powers of $2W/cm^2$, $0.5W/cm^2$ and $0.25W/cm^2$ utilized[19]. It sould be mentioned here for avoiding side effects we use more accurate laser with less powers and in real experiments one laser can be used several times. The laser is countinious laser with the exposure time of 50 s and cooling time of 20 s. Table 3 shows the optical poperties of the tissue containing the tumor. Moreover, for controlling the produced temperature by ultrasound, it radiate to tumor for 60s and then it turn off for 10s. This process continues for 10 min.

In addition, for increasing the heating absorption, gold nanoparticles are added to the medium. The nanoparticles are assumed to be homogeneously destributed in the tissue. In this simulation, the absorption and scattering coeffient of nanoparticles are 121/cm and 5/cm, respectively[21].

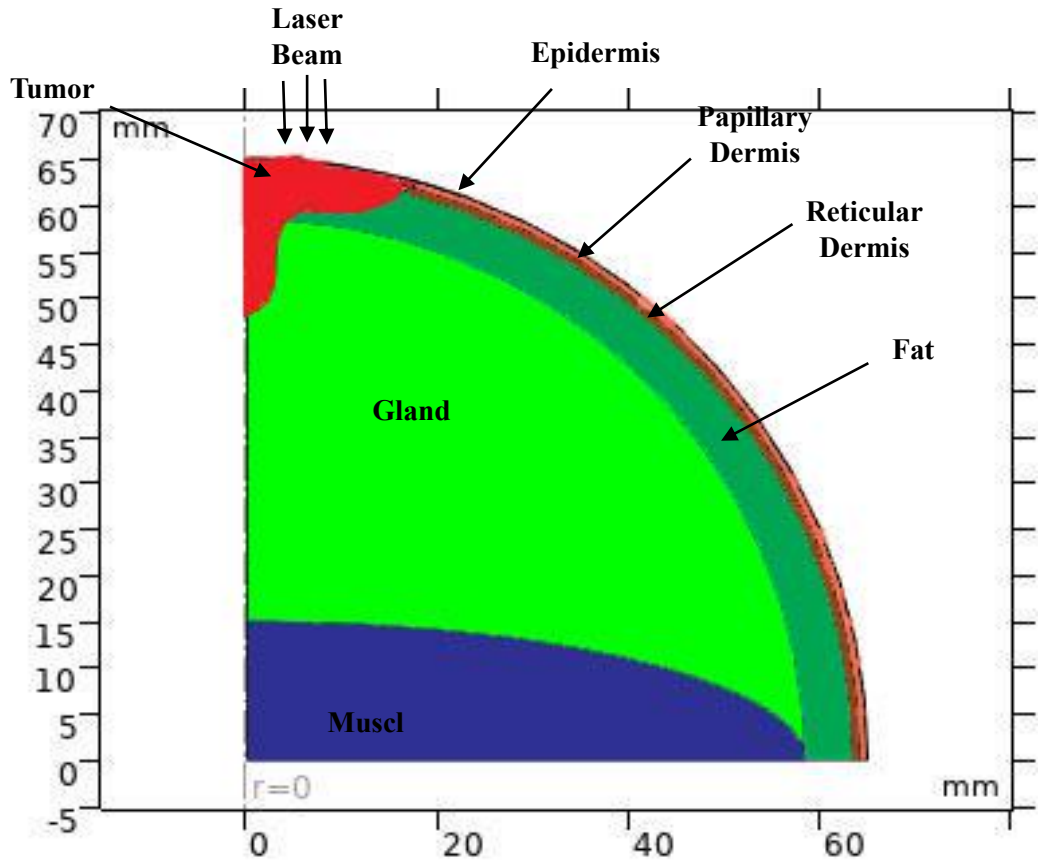

Figure 1. The geometry of the breast tissue and the tumor.

Table1. Physical and thermal properties of tissue and tumor[3].

| Layer | Epidermis | Papillary Dermis | Reticular dermis | Fat | Gland | Muscle |
|---|---|---|---|---|---|---|
| h(mm) | 0.1 | 0.7 | 0.8 | 5 | 43.4 | 15 |
| k (W/ (m.K)) | 0.235 | 0.445 | 0.445 | 0.21 | 0.48 | 0.48 |
| $\rho$ (kg/m$^3$) | 1200 | 1200 | 1200 | 930 | 1050 | 1100 |
| c (J/ (kg. K)) | 3589 | 3300 | 3300 | 2770 | 3770 | 3800 |
| $Q_0$(W/m$^3$) | 0 | 368.1 | 368.1 | 400 | 700 | 700 |
| $\omega_b$ (1/s) | 0 | 0.0002 | 0.0013 | 0.0002 | 0.0006 | 0.0009 |

Table 2. Thermal properties of blood[22].

| property | value |
|---|---|
| $C_b (J/(kg.K))$ | 4200 |
| $\rho$ (kg/m^3) | 1000 |
| $T_b(K)$ | 310.15 |

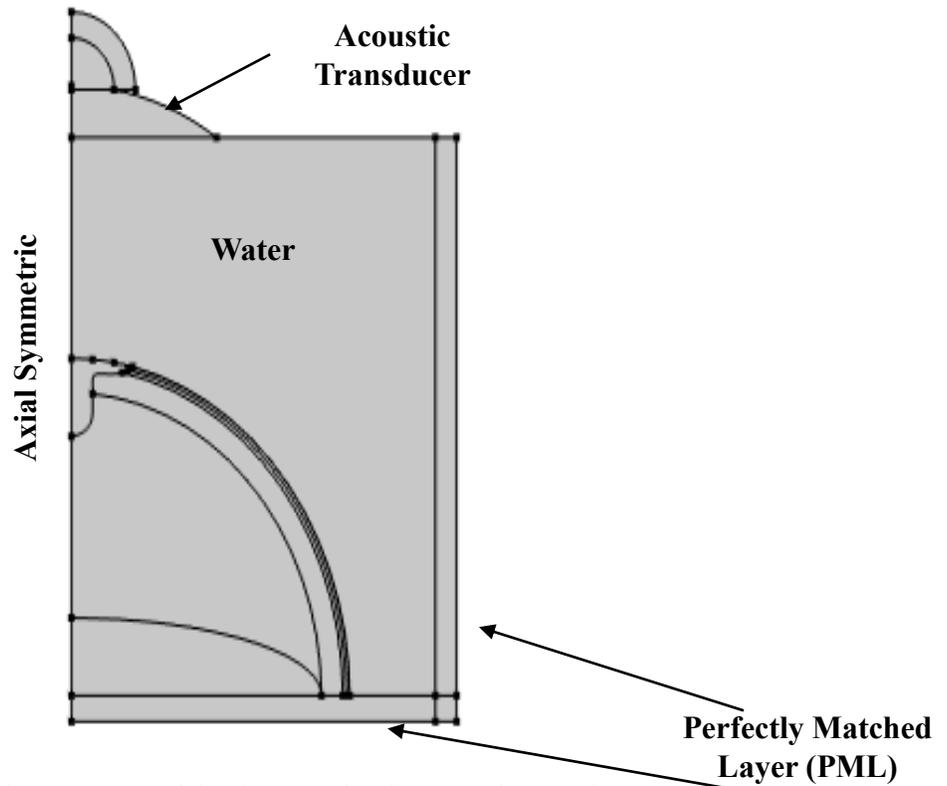

Figure 2. The geometry of the tissue under the acoustic transducer.

Table3. The optical characteristics of the breast tissue[23,24].

| Layer | Epidermis | Papillary Dermis | Reticular dermis | Fat | Gland | Tumor |
|---|---|---|---|---|---|---|
| $\mu_a$(1/m) | 19 | 19 | 19 | 6.5 | 6 | 6 |
| $\mu_s$(1/m) | 834 | 834 | 834 | 1000 | 1275 | 5 |
| $\alpha$(dB/m) | 35 | 35 | 35 | 48 | 75 | 43 |
| C(m/s) | 1510 | 1510 | 1510 | 1510 | 1510 | 1540 |

## 2.2 bioheat transfer

Thermal cancer treatment plays a significant role in the field of the clinical applications. This kind of treatment causes alteration of the tissue temperature to achieve the thermal ablation of tumors. In this procedure, an ultrasound transducer, a laser beam, or microwave applies mechanical or light energy to tissues, leading to an elevation in temperature and subsequently, causing the demise of cancerous cells[25]. The issue with the physical bioheat problems, is evaluation of the temperature profile and the fluence rate distribution in the tumorous region with knowing the governing initial and boundary conditions, geometry, heat sources, as well as the thermo physical and optical properties of the tissues. Generally, in living tissues, blood perfusion and passage of blood modifies the heat transfer. Furthermore, metabolic activity generates heat within the tissue. Therefore, an equation is needed for describing the heat transfer in tissue by considering the effects of both blood perfusion and metabolism.

This relation, widely known as Penne's equation, was first established by Penne (1948) and Perl (1962)[26,27,28] as below:

$$\rho c_P \frac{\partial T}{\partial t} = \nabla \cdot (k\nabla) - \rho_b \omega_b C_b (T - T_b) + Q_{met} + Q_s \tag{1}$$

where, $\rho$ is density of tissue, $c_P$ is heat capacity, T is temperature, t is time, $k$ is thermal conductivity of tissue, $\rho_b$ is blood density, $\omega_b$ is blood perfusion, $c_b$ is heat capacity of blood, $T_b$ is temperature of blood, $Q_{met}$ indicates the metabolic heat source term, and $Q_s$ is the external heat source. Here, there are two heat sources, laser and ultrasound as follow:

$$O_s = Q_L + Q_{US} \tag{2}$$

In the following sections, each of these heat sources will be analyzed and then inserted into Eq. (1)

## 2.3 Interaction of laser and tissue

In this research, laser beam is one of the main heat sources for tumor ablation. When an incident laser beam enters the medium in direction $\Omega$, part of its intensity is absorbed and another part of that is scattered in another direction, $\Omega'$. The fraction that is absorbed is shown by $\sigma_a$ I($\Omega$) and the scattered fraction is indicated by $\sigma_s$ I($\Omega$), where $\sigma_a$ and $\sigma_s$ are absorption and scattering coefficients. For obtaining the energy deposited at each point, one needs to solve the radiative transfer equation (RTE). For quasi-static conditions, which are met in the current study, RTE obeys the following equation[29]:

$$\Omega \cdot \nabla I(\Omega) = \sigma_a I_b(T) - (\sigma_a + \sigma_s)I(\Omega) + \frac{\sigma_s}{4\pi} \int_{4\pi} I(\Omega')\phi(\Omega', \Omega) d\Omega' \tag{3}$$

In the above equation, $I_b$ denotes the blackbody intensity, which is derived from the Plank function and $\phi$ ($\Omega'$, $\Omega$) signifies the probability that the beam will scatter from direction $\Omega'$ toward the direction $\Omega$. The last term of the right-hand side of this equation accounts for the portion of the radiative energy coming from all possible directions which is scattered toward the considered direction of propagation.

Generally, different methods such as discrete ordinates method (DOM), P1-approximation, Rosseland approximation, and Beer–Lambert law are employed to solve the RTE. The P1 approximation stands out as one of the best and the most straightforward methods for solving the RTE, utilizing spherical harmonics with the capability of incorporating both isotropic and linear anisotropic scattering within its framework. In P1-approximation method, incident radiation is defined as[30]:

$$G = \int_{4\pi} I(\Omega) \, d\Omega \qquad (4)$$

and, radiative heat flux is expressed as[30]:

$$q_r = \int_{4\pi} I(\Omega) \, d\Omega \qquad (5)$$

By applying G and $q_r$ to the RTE, it can be rewritten as below[30]:

$$I(r, \widehat{\Omega}) = \frac{1}{4\pi}[G(r) + 3q(r) \cdot \widehat{\Omega}] \qquad (6)$$

When the laser beam irradiates into the medium, I in the equation will be consisted of the collimated laser beam entering the medium and the diffusion term emerging from scattering within the medium. As described by Modest and Tabanfar in Ref.[30], in such scenarios, the impact of the external radiation field can be taken into account by introducing a source term into the diffusion terms. Hence, the simplified version of RTE emphasizes on the diffusion of light due to scattering phenomenon, and the source term which directly incorporates the influence of the external collimated beam from the laser as bellow[30]:

$$\nabla \cdot (D_{P1} \nabla G) - \sigma_a(G - 4\pi I_b) = 0 \qquad (7)$$

here, $D_{P1}$ is diffusion coefficient in P1-approximation method and is defined as[30]:

$$D_{P1} = \frac{1}{3\sigma_a + \sigma_{s(3-a_1)}} \qquad (8)$$

where, $a_1$ is the linear Legendre coefficient for the scattering phase function. This coefficient is related to anisotropy so that for the case of isotropic scattering a1 equals to zero.

Then, in the Penne equation, $Q_L$ is the laser heat source which is calculated by the divergence of heat flux $q_r$. That is because the radiative heat flux crossing an element with an area normal to the direction of $\boldsymbol{\Omega}$ is a result of intensities incident from all directions and it is the divergence of heat flux at any point which leads to rising the temperature at that point. It should be stressed that once G is characterized at each point, $\nabla \cdot q_r$ can be found.

Moreover, addition of gold nanoparticles significantly modifies the optical features of the tissue. In this paper absorption and scattering coefficient of gold nanoparticles are 121/cm and 0.5/cm, respectively[21].

## 2.4 Interaction of ultrasound with tissue

As previously mentioned, ultrasound can be considered as a heat source. Here, the absorbed acoustic energy is calculated and used as the heat source, $Q_{US}$, in Penne's equation.

The propagation of the focused ultrasound waves as a time-independent problem follows the homogenous Helmholtz equation in 2D axisymmetric cylindrical coordinates as below[20]:

$$\frac{\partial}{\partial r}\left[-\frac{r}{\rho_c}\left(\frac{\partial P}{\partial r}\right)\right] + r\frac{\partial}{\partial z}\left[-\frac{1}{\rho_c}\right] - \left[\left(\frac{\omega}{C_c}\right)^2\right]\frac{rp}{\rho_c} = 0 \qquad (9)$$

where, r and z are the radial and axial coordinates, p is the acoustic pressure, and $\omega$ is the angular frequency. $\rho_c$ is density and $C_c$ is the speed of sound which are also complex-valued to account for the material's damping properties.

In Helmholtz equation, it is assumed that the acoustic wave propagation is linear and the amplitude of the shear waves in the tissue domain are much smaller than that of the pressure waves. Therefore, nonlinear effects and shear waves are neglected.

The acoustic intensity field can also be derived from the acoustic pressure field. Finally, by knowing the acoustic intensity field, $Q_{US}$ is calculated as below[20]:

$$Q_{US} = 2\alpha_{ABC} I = 2\alpha_{ABC} \left| \text{Re}\left(\frac{1}{2}pv\right) \right| \qquad (10)$$

here, $\alpha_{ABC}$, I, p and v are attenuation coeffient, acoustic intensity magnitude, the acoustic pressure, and acoustic particle velocity vector, respectively. Furthermore, it should be mentioned that attenuation coefficient of tissue with and without gold nanoparticle is 70 dB/m and 43 dB/m, respectively[24].

## 2.5 Thermal damage

An important concern in research about bioheat transfer is accurate and appropriate thermal damage of the diseased tissues without any destruction of the neighboring healthy tissues during tumor treatment. Several investigations have demonstrated that tissue damage depends on both of the temperature and exposure time. It should be mentioned that as tissue temperature increases, the elapsed time necessary to achieve the threshold of damages decreases. The progression of the thermal injury can be reasonably approximated by Arrhenius equation. This relation is usually utilized for illustrating the rate of the irreversible heating damage of the biological tissues[31,32]:

$$\Omega(t) = \int_0^t A e^{\frac{\Delta E}{RT}} dt \tag{11}$$

In the above equation, $\Omega(t)$ indicates the degree of tissue injury, R is the universal gas constant, A is the frequency factor for kinetic expression, and ¬E is the activation energy for irreversible damage reaction. The critical value of $\Omega = 1$ is the point when thermal necrosis occurs. Generally, two parameters of A and $\Delta E$ depend on the type of tissue. Here, for the breast tissue A and $\Delta E$ are $1.18 \times 10^{44}$ and $3.02 \times 10^{55}$, respectively[3].

## 3. Validation of the simulation results

To validate the accuracy of the proposed model, the results of the ultrasound transducer simulation and nanoparticle-mediated laser therapy model are compared with previous literatures.

It is important to mention that the simulation of laser photothermal therapy with nanoparticles in the current model is based on the same model utilized in our previous paper[19] on liver cancer treatment. In our previous paper, it was shown that under similar conditions, the results of our simulations agree well with those obtained by Sony et al[21] [see figure 3 of our previous article[19]].

In order to validate the results of the FEM simulation of the ultrasonic probe, the present model is compared to the research conducted by Huang et al.[33] in 2004. Figure 3 depicts the 2D axisymmetric geometry used by Huang et al. in their paper. For validation, the same model with similar dimensions, boundary conditions, and physical properties is considered to measure the temperature of the tissue phantom at the ultrasound focus. Figure 4 illustrates the heating and cooling curves characteristic of this study, along with the results from the experiments and simulations conducted by Huang et al. As it is clearly seen in this figure, the data of the present model is in complete agreement with the results of the simulations of the previous literature. Nonetheless, there is a small difference between the experimental data and the data of the current simulation. These differences can be attributed to some approximations assumed in the simulation, while the overall fit to the simulation curve is quite remarkable.

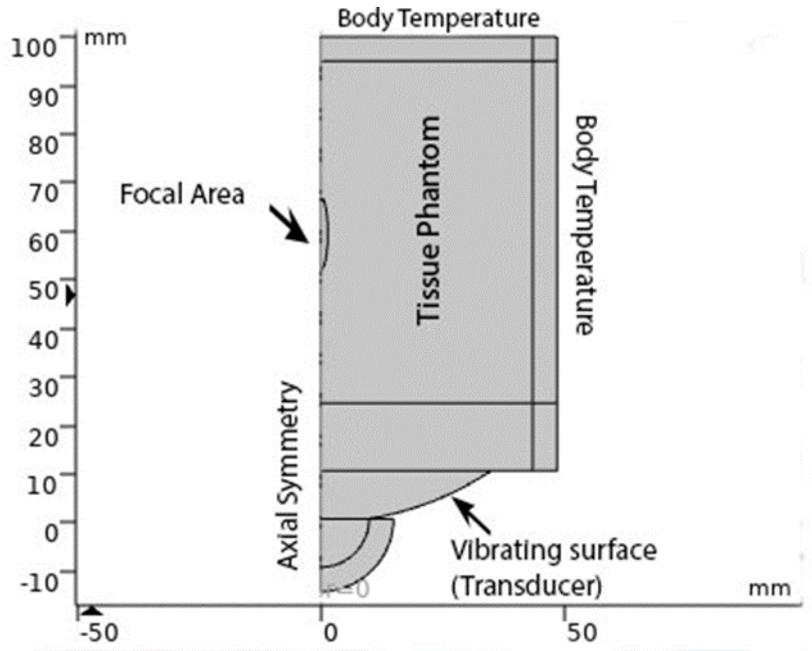

Figure 3. The geometry of the model produced by Huang et al.[33], which is used for validation of the data of focused ultrasound simulation.

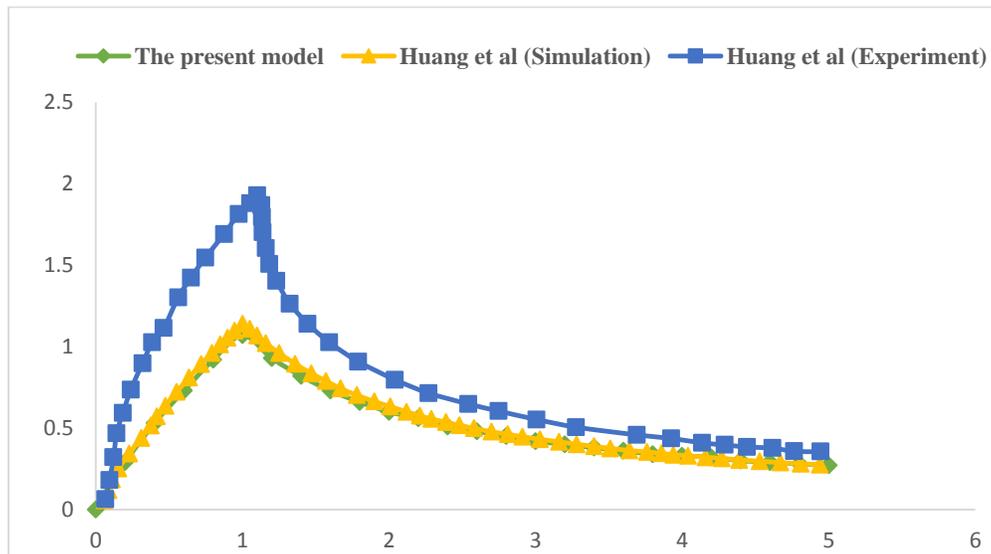

Figure 4: Comparison between the focal temperatures calculated by Huang et al.[33] and the present model.

## 4. Results and Discussion

This study utilized numerical analysis to examine the tissue's mechanical and thermal response to ultrasound and laser exposure in order to determine how adjustable parameters can precisely ablate a tumor, while minimizing nociceptive pain and causing no damage to adjacent healthy tissue. Usually, it is difficult to remove the tumor without causing harm to the healthy tissues around. Therefore, to reduce the possibility of thermal tissue damage, gold nanoparticles are introduced into the tumor when it is being exposed to diode laser irradiation. This raises the tumor's temperature above normal, while shielding the surrounding healthy tissues. It should be noted that the lasers used in this instance are pulsed lasers, meaning they heat up for 50 seconds and cool down for 20 seconds. The problem is performed twice, once with and once without the insertion of Au nanoparticles, in order to determine the impacts of gold nanoparticles on the enhancement of temperature in the tissue. The findings of the comparison are then displayed.

## 4.1 The effect of Nanoparticles on heating by laser and ultrasound

To understand the effect of nanoparticles on increasing the temperature of tumor region and the amount of necrosis, the problem is solved once with nanoparticles and again without nanoparticles. The laser used in this study is a continuous laser with the exposure time of 50s and cooling time of 20s. Also the exposure time of US is 60s and its cooling time is 10s. Figures 5 shows the temperature of the tissue after laser irradiation, with and without presence of gold nanoparticles for 10 min. In the absence of gold nanoparticles, the temperature of the tumor reaches 38°C and when gold nanoparticle is injected, the temperature increases to 54°C. And the figure 6 shows the volume of necrosis without and again with gold nanoparticles injection to the tumor after 10 min of US irradiation. As it can be seen in the picture 6, injection of gold nanoparticle causes the cell death in deep section of the tumor noticeably increase. As well as the depiction of the acoustic pressure in the water and tissue domain is given in figure 7.

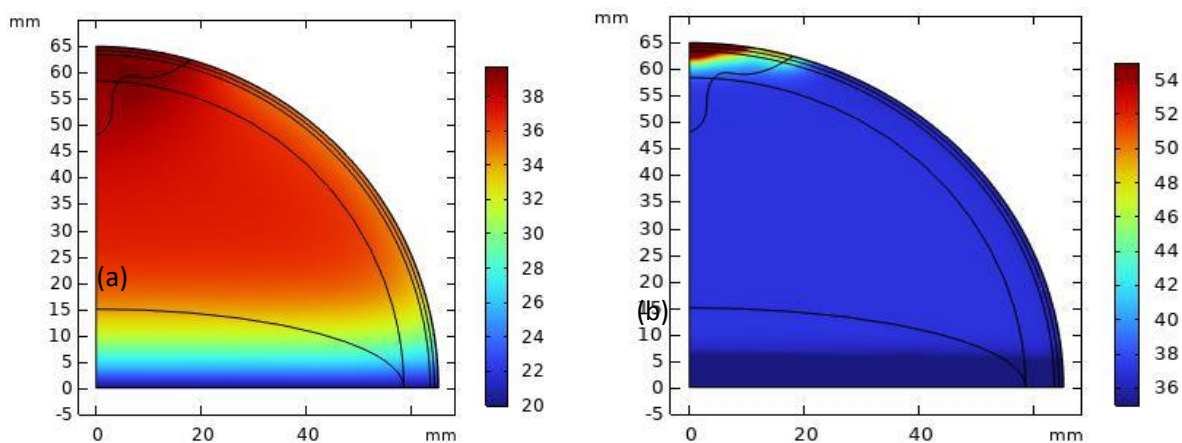

Figure 5. Temperature of the tissue after 10 min of laser irradiation (a) without, and (b) with gold nanoparticles.

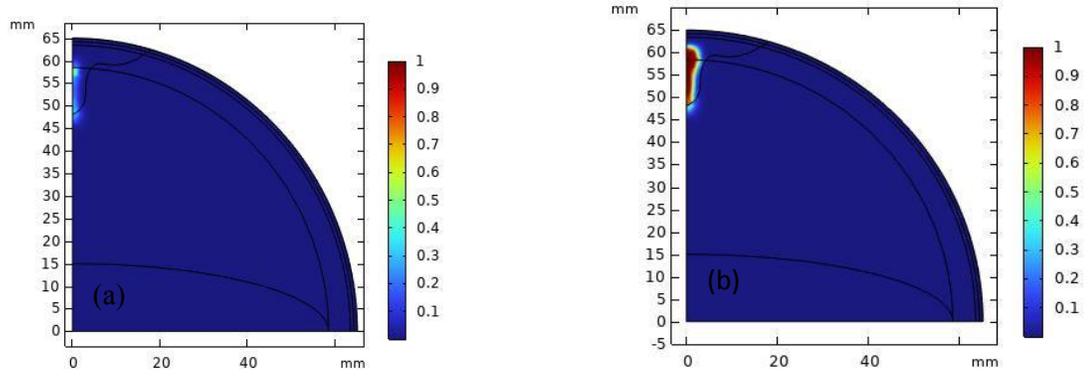

Figure 6. the faction of necrosis after 10 min of ultrasound irradiation (a) without, and (b) with gold nanoparticles injection.

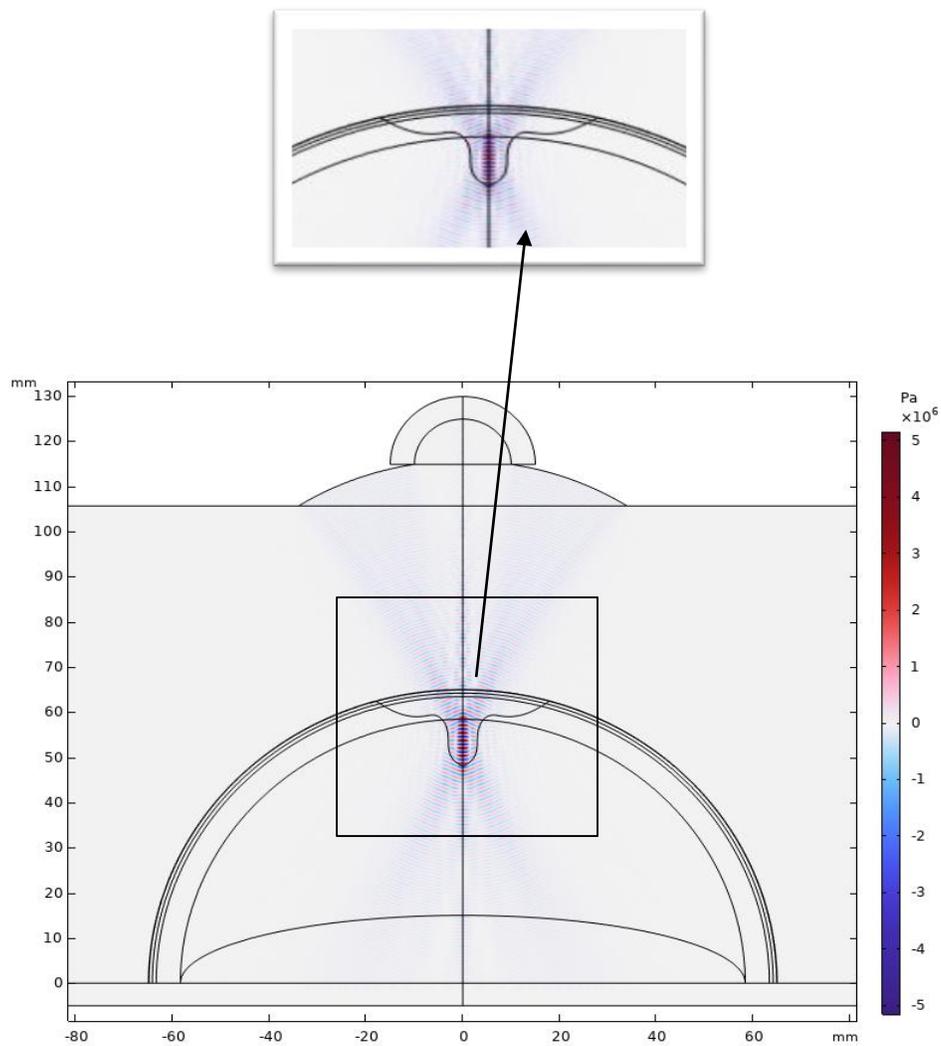

Figure 7. The acoustic pressure in the water and tissue domain.

## 4.2 Combining focused ultrasound and laser photo thermal therapy

Figure 6 shows the temperature rise in the tissue after 10 min of receiving laser and ultrasound irradiation, simultaneously. As it can be seen in this figure 6, the minimum temperature is 20°C, i.e. the body temperature, and the maximum temperature reaches 100°C after 10 minutes of laser and ultrasound irradiation together.

The important issue that determines the successful of method is the fraction of damaged tissue, which is presented in figure 7. Figure 7a shows the fraction of the damage after 10 min of laser illumination, while figure 7b presents the damage fraction after 10 min ultrasound irradiation. Finally, figure 7c illustrates the status when the tissue is simultaneously irradiated by both of laser and ultrasound for 10 min.

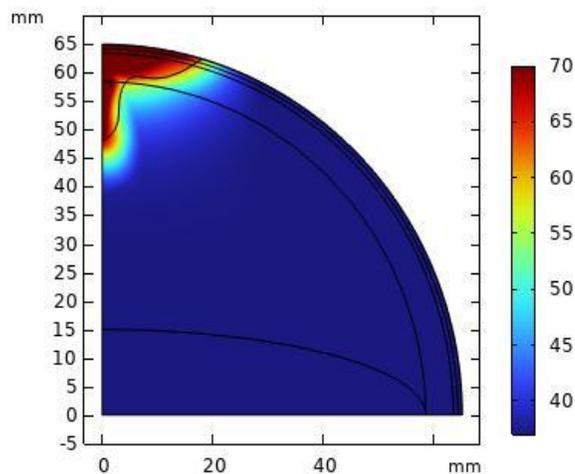

**Figure 6**. Temperature evolution in the tissue after 10 min of receiving laser and ultrasound irradiation simultaneously.

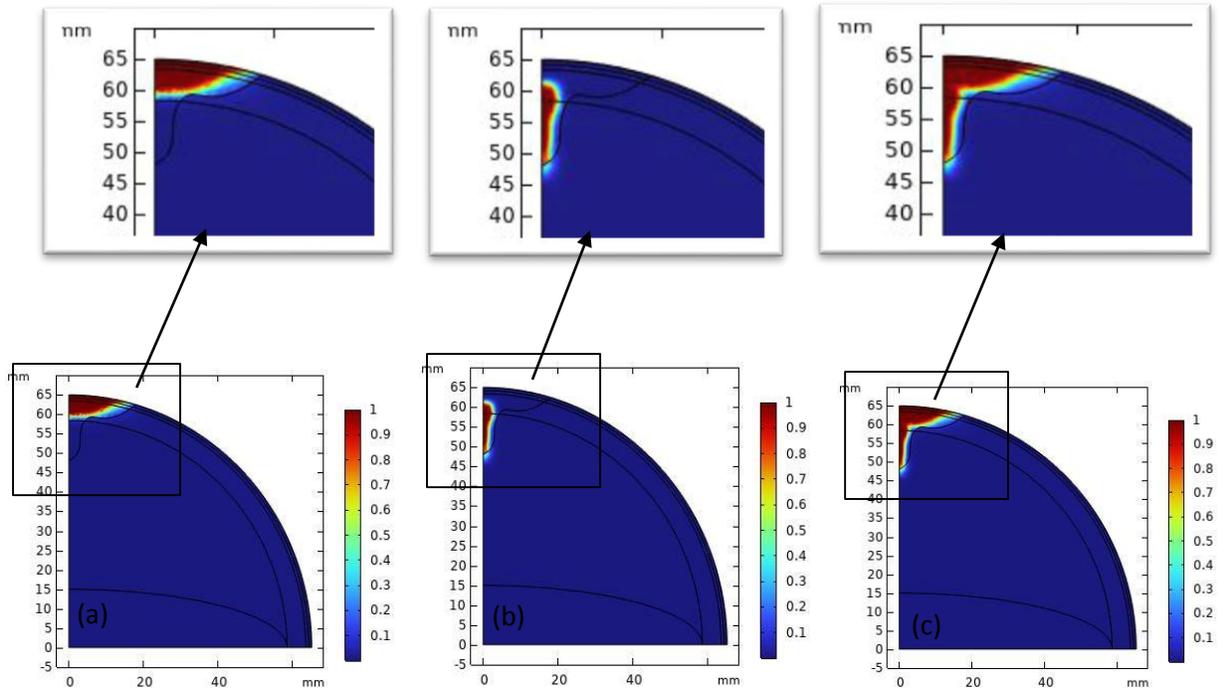

**Figure 7.** The fraction of damaged tissue after (a) 10 min of laser illumination, (b) 10 min irradiation of ultrasound, and (c) 10 min of irradiation of both of laser and ultrasound.

To check that this method has not damaged the surrounding healthy tissue, the fraction of the necrotic tissue along four lines, crossing the tissue model is evaluated at different places. For this purpose, two vertical and two horizontal lines are chosen to help us to have a better judgment on the damage estimation in both superficial and deep areas. The horizontal lines are positioned at z = 54 and z = 60, while the vertical lines are located at r = 6 and r = 10 of figure 8. The necrosis fraction along the horizontal and vertical lines are shown in figures 9 and 10, respectively.

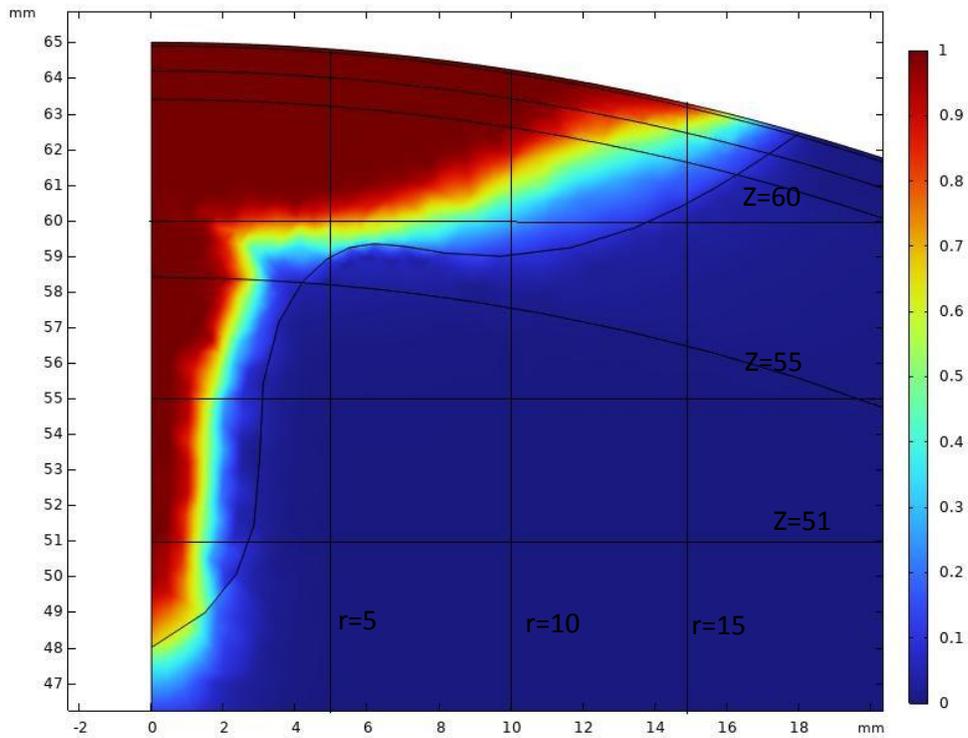

**Figure 8.** Four horizontal and vertical lines for assessment of the amount of damage to the neighboring areas of tumor.

According to figure 8, the border of the safe region is shown by the yellow color (i.e., necrosis fraction equal to 0.7), indicating that there is no serious damage present in the normal sections of tissues and the red area indicates damage within the tumor. Therefore, it is obvious that this method is both safe and effective for treating cancers.

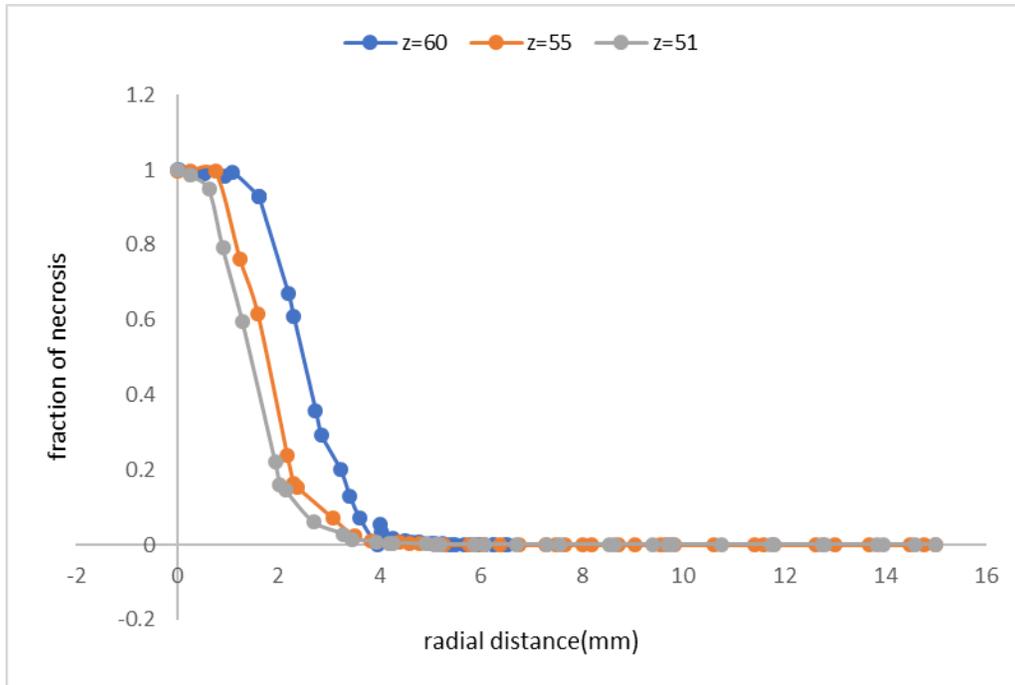

**Figure 9.** The fraction of tissue's necrosis versus radial distances at constant horizontal positions of z = 50, 55 and 60 mm.

In addition, as seen in figure 9, at z=60 mm and at reduced radial distances, wider areas are more vulnerable to necrosis because they are near to laser and ultrasound radiations. Moreover, the necrosis proportion is minimal at long radial distances in all horizontal lines, demonstrating the lack of adverse effects from this treatment approach.

A similar pattern is seen in figure 10 at various radial distances, with a slight increase at r = 6 mm. Furthermore, as this figure makes evident, the necrosis fraction is greater than 0.7 at z > 62 mm, indicating that only the tumor areas are damaged and supporting the usefulness of the suggested method.

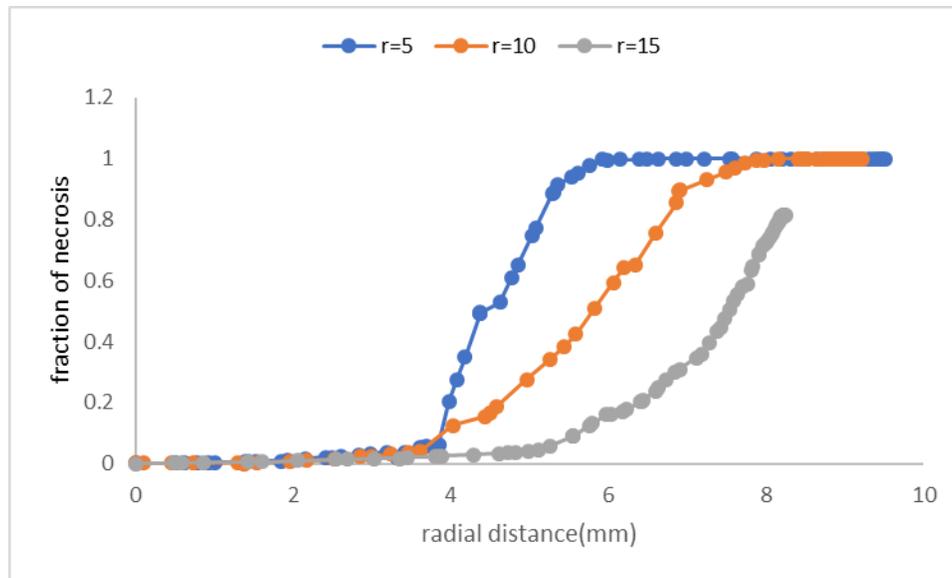

**Figure 10.** The fraction of tissue's necrosis along different horizontal lines, at fixed radial distances of r = 6, 10 and 15 mm.

## Conclusion

In this research, the interaction between laser and breast tissue in the presence of gold nanoparticles combined with ultrasound irradiation is studied. The FEM (finite element method) simulations revealed that this treatment approach could be exceptionally beneficial for treatment of the tumors including both superficial and deep regions. Here, the heat generated by the acoustic transducer was estimated using the acoustic intensity magnitude. Furthermore, the magnitude of energy produced by laser irradiation and absorbed by GNPs was determined by solving the RTE equation. For solving RTE equation, P1-approximation was employed which means isotropic and linear anisotropic scatterings were both considered. Finally, Pennes' equation was solved to calculate the temperature within the tumor.

In summary, the results showed that: 1) Gold nanoparticle injection greatly enhanced the effectiveness of the photothermal therapy. 2) In the absence of gold nanoparticles, the temperature raised just into 38°C after 10 min laser irradiation. 3) With the injection of gold nanoparticles, the temperature of laser irradiation increased up to 100°C after 10 min of laser irradiation. 4) Moreover, gold nanoparticles caused the temperature rise up to 100°C after ultrasound irradiation. 5) The effect of gold nanoparticles on enhancing the heating effect of laser was more than ultrasound irradiation, and 6) finally, after 10 min irradiation of laser and ultrasound, most parts of the tumor undergo necrosis. Hence, it can be concluded the proposed irradiation of laser and ultrasound with the aid of gold nanoparticles could help in treatment of cancer with no damage to healthy tissue.

**Author contributions**